\documentstyle[12pt]{article}

\newcommand{\pabar}{\not{\!\partial}}
\newcommand{\Od}{{\cal O}}

\newcommand{\tr}{\mbox{tr}}

\newcommand{\Dbar}{\not{\!{\!D}}}

\begin{document}
\input epsf \renewcommand{\topfraction}{0.8}
\pagestyle{empty} \vspace*{5mm}
\begin{center}
\Large{\bf Goldstone bosons and solitons on the brane$^*$}
\\ \vspace*{2cm}
\large{ J. A. R. Cembranos, A. Dobado and A. L. Maroto}
\\ \vspace{0.2cm} \normalsize
Departamento de  F\'{\i}sica Te\'orica,\\
 Universidad Complutense de
  Madrid, 28040 Madrid, Spain\\
\vspace*{0.6cm} {\bf ABSTRACT} \\
\end{center} \vspace*{5mm}
In this work we compute the effective action describing
the low-energy brane Goldstone-boson dynamics. The
Goldstone bosons or branons appear because the brane
spontaneously breaks the translational invariance along
the extra dimensions. We discuss the Higgs-like mechanism in
which the Kaluza-Klein gauge fields absorb the Goldstone bosons
and acquire mass. We also present the explicit form of the
effective action describing the low-energy
 interactions between the three-brane Goldstone bosons and the
Standard Model fields. In addition to the branons, two new kinds of states
arise corresponding to Skyrmion-like solutions and
wrapped states. We study the main properties of these
states, their geometrical interpretation and the relations
between them. 

\vspace{1cm}

\noindent
\rule[.1in]{8cm}{.002in}

\noindent
$^*$Contributed to 20th International Symposium on Lepton and 
Photon Interactions at High Energies (Lepton Photon
01), Rome, Italy, 23-28 July 2001

\newpage
\setcounter{page}{1} \pagestyle{plain}

\textheight 20 true cm
\section{Introduction}
The possibility of having  large extra dimensions has been considered
in recent works
as a  solution of the
hierarchy problem. In this proposal, the most relevant scale would be the
Planck scale in $D=4+n$ dimensions $M_D$, which is assumed to be
not too far from the TeV scale. In the original approach \cite{Hamed1},  
the much larger value of the
4-dimensional Planck constant would be due to the size of the
extra dimensions, since we typically have $M^2_P \simeq
R^nM_D^{n+2}$, where $R$ is the  radius of the compactified extra
dimensions. Phenomenological considerations then require that,
whereas gravity can propagate in the D-dimensional bulk space, the
ordinary matter fields and gauge bosons be bound to live on
a 3-dimensional brane,  which would constitute
the usual spatial dimensions. The brane tension $f$ is the other
important scale in this scenario, $f^{-1}$ being the typical size
of the brane fluctuations. 

Special attention has been
paid to the description of the low-energy sector of the model.
This sector would include the Standard Model (SM) fields, the
gravitons and the possible excitations of the brane
\cite{Sundrum}. The presence of  extra dimensions allows for
the existence, in addition to the standard massless gravitons in
four dimensions, of an infinite tower of massive gravitons
(Kaluza--Klein \cite{KK} gravitons) whose mass is determined by the
size of the extra dimensions (see for example \cite{Balo} for a
review of different Kaluza--Klein models). The interaction of the
graviton sector with the SM fields has been analysed in a series
of papers \cite{Giudice} and different predictions have been
obtained that could be tested at future particle colliders.
However, in addition to the gravitons, the low-energy spectrum of
the theory also contains the brane's own excitations (branons). 
If the brane
has been spontaneously created with a given shape in the bulk (that
we will consider as its ground state), the initial isometries of
the bulk space could be broken spontaneously by the presence of
the brane. The brane configurations obtained by means of some
isometry transformations in the bulk will be considered as
equivalent ground states and therefore the parameters describing
such transformations can be considered as zero-mode excitations of
the ground state. When such transformations are made local
(depending on the position on the brane), the corresponding
parameters play the role of Goldstone bosons (GB) fields of the
isometry breaking. Moreover, it has been shown that, in the case
where the brane tension $f$ is much smaller than the fundamental
scale $M_D$ $(f\ll M_D)$, the non-zero KK modes decouple from the
GB modes \cite{GB,Kugo} and it is then possible, at least in principle,
to make a low-energy effective theory description of the GB
dynamics. On the other hand, in the standard Kaluza--Klein models,
the isometries in
the extra dimensions are understood as gauge transformations in
the four-dimensional theory. Therefore, since the GB are
associated to the breaking of those {\it gauge} transformations,
it is natural to expect \cite{Hamed2} that some kind of Higgs
mechanism can take place, which would  give mass to the
Kaluza--Klein gauge bosons.

In addition to the branons, the brane can support also a new set of
topological states. These states are defects that appear due to
the non-trivial homotopies of the vacuum manifold \cite{Shifman}.
In particular the authors of this reference have considered the
case of string and monopole defects on the brane, corresponding
to non-trivial first and second homotopy groups.

In this work we are interested in another kind of defects of
topological nature  related to the Skyrme model \cite{Skyrme}. In
this model, the baryons are understood as topological solitons
that appear in the low-energy pion dynamics described by a chiral
lagrangian, the baryon number being identified with the
topological charge. This model has provided a very successful
description of the baryon properties \cite{Witten2}. The
branon effective action mentioned before 
is formally similar \cite{Dobado} 
to the chiral lagrangians used for the low-energy
description of the chiral dynamics \cite{Weinberg} or even the
symmetry breaking sector of the Standard Model in the strongly
coupled case \cite{DH}. Therefore it is quite natural to wonder about the
possibility of having chiral solitons (Skyrmions) arising from
this effective action. As we will show the answer is positive. In
the following we will study in detail those brane-skyrmions, their
physical and geometrical interpretation, their main properties and
their relation with wrapped states.

The plan of the paper goes as follows: In Sec.2 we introduce our
set up and obtain the brane GB effective
action starting from a generalized brane action that includes an
induced scalar curvature term. This term will be essential in
order to determine the brane-skyrmion size. In Sec.3, we describe the
Higgs-like mechanism we have commented before.
In Sec.4 we derive the coupling of branons with the SM fields. Sec.5 
is devoted to the equations for the
brane-skyrmions and there we also compute analytically 
their size and mass in
terms of the different parameters involved.
In Sec.6 we consider the
interacions between the brane-skyrmions and branons and  the
possible fermionic quantization. In Sec.7 we
consider another  set of brane states (wrapped states) and study
their relation with the brane-skyrmions. Finally, in Sec.8 we set
the main results of our work and the conclusions.

\section{The effective action for the branons}

Let us start by fixing the notation and the main assumptions used
in the work.  We consider that the four-dimensional space-time
$M_4$ is embedded in a $D$-dimensional bulk space that for
simplicity we will assume to be of the form $M_D=M_4\times B$,
where $B$ is a given N-dimensional compact manifold so that
$D=4+N$. The brane lies along $M_4$ and we neglect its
contribution to the bulk gravitational field. The coordinates
parametrizing the points in $M_D$ will be denoted by
$(x^{\mu},y^m)$, where the different indices run as $\mu=0,1,2,3$
and $m=1,2,...,N$. The bulk space $M_D$ is endowed with a metric
tensor that we will denote by $G_{MN}$, with signature
$(+,-,-...-,-)$. For simplicity, we will consider the following
ansatz:
\begin{eqnarray}
 G_{MN}&=&
\left(
\begin{array}{cccc}
\tilde g_{\mu\nu}(x)&0\\ 0&-\tilde g'_{mn}(y)
\end{array}\right).
\end{eqnarray}
In the absence of the 3-brane, this metric possesses an isometry
group that we will assume to be of the form $G(M_D)=G(M_4)\times
G(B)$. The presence of the brane spontaneously breaks this
symmetry down to some subgroup $G(M_4)\times H$. Therefore, we can
introduce the coset space $K=G(M_D)/(G(M_4)\times H) =G(B)/H$,
where $H\subset G(B)$ is a suitable subgroup of $G(B)$.

The position of the brane in the bulk can be parametrized as
$Y^M=(x^\mu, Y^m(x))$, where we have chosen the bulk coordinates
so that the first four are identified with the space-time brane
coordinates $x^\mu$. We assume the brane to be created at a
certain point in $B$, i.e. $Y^m(x)=Y^m_0$ which corresponds to its
ground state. The induced metric on the brane in such state is
given by the four-dimensional components of the bulk space metric,
i.e. $g_{\mu\nu}=\tilde g_{\mu\nu}=G_{\mu\nu}$. However, when
brane excitations (branons) are present, the induced metric is
given by
\begin{eqnarray}
g_{\mu\nu}=\partial_\mu Y^M\partial_\nu Y^N G_{MN} =\tilde
g_{\mu\nu}-\partial_{\mu}Y^m\partial_{\nu}Y^n\tilde g'_{mn}.
\end{eqnarray}
For illustrative purposes we show a toy model in Fig. 1 where we
have a 1-brane (string) in a $M_3=M_2\times S^1$ bulk-space, both
in its ground state (flat brane) and in an excited state (wavy
brane).

Since the mechanism responsible for the creation of the brane is
in principle unknown, we will assume that the brane dynamics can
be described by  an effective action. Thus, we will consider the
most general expression which is invariant under
reparametrizations of the brane coordinates. Following the
philosophy of the effective lagrangian technique, we will also
organize the action as a series in the number of the derivatives
of the induced metric over a dimensional constant, which can be
identified with the brane tension scale $f$. Therefore, up to
second order in derivatives we find:
\begin{equation}
S_B=\int_{M_4}d^4x\sqrt{g}\left( -f^4+\lambda f^2 R + \dots\right),
\label{Nambu4}
\end{equation}
where $d^4x\sqrt{g}$ is the volume element of the brane, $R$ the
induced curvature and $\lambda$ an unknown dimensionless
parameter. Notice that the lowest order term is the usual
Dirac-Nambu-Goto action that was the only one considered in
\cite{Dobado}.
\begin{figure}
\vspace*{0cm}
\centerline{\mbox{\epsfysize=5.5 cm\epsfxsize=5.5
cm\epsfbox{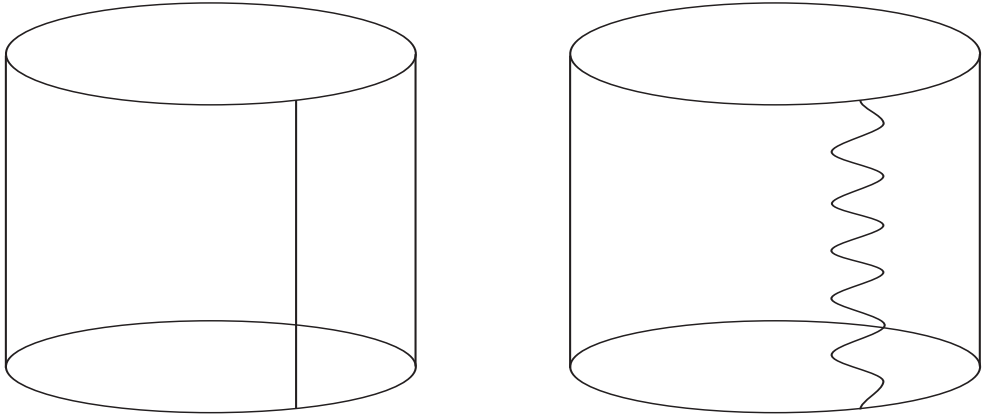}}\hspace*{2 cm}\mbox{\epsfysize=5.5
cm\epsfxsize=5.5 cm\epsfbox{cilmonel1.eps}}}
%
 \caption{\footnotesize{Brane with trivial topology in $M_3=M_2\times S^1$.
The ground
state of the brane is represented on the left. On the right we plot
an excited stated.}}
\end{figure}

If the brane ground state is $Y^m(x)=Y^m_0$, the presence of the
brane will break spontaneously all the $B$ isometries, except
those that leave the point $Y_0$ unchanged. In other words the
group $G(B)$ is spontaneously broken down to $H(Y_0)$, where
$H(Y_0)$ denotes the isotropy group (or little group) of the point
$Y_0$. Let $H_i$ be the $H$ generators ($i=1,2,...\; h$),
$X_\alpha$ ($\alpha=1,2,...\;k=\mbox{dim}\; G(B)-\mbox{dim}\;H$)
the broken generators, and $T=(H,X)$ the complete set of
generators of $G(B)$. A similar separation can be done with the
Killing fields. We will denote $\xi_i$ those associated to the
$H_i$ generators, $\xi_\alpha$ those corresponding to $X_\alpha$
and by $\xi_a(y)$
 the complete set of Killing
vectors on $B$. The excitations of the brane along the (broken)
Killing fields directions of $B$ correspond to the zero modes and
they are parametrized by the branon fields $\pi^\alpha(x)$ that
can be understood as coordinates on the coset manifold $K=G(B)/H$.
Thus, for a position-independent ground state $Y^m_0$, the action
of an element of $G(B)$ on $B$ will map $Y_0$ into some other
point with coordinates:
\begin{equation}
Y^m(x)=Y^m(Y_0,\pi^\alpha(x))=Y^m_0+\frac{1}{k
f^2}\xi^m_\alpha(Y_0)\pi^\alpha(x)+\Od(\pi^2), \label{ypi1}
\end{equation}
where we have set the appropriate normalization of the branon
fields and Killing fields   with $k^2=16\pi /M_P^2$. It is
important to note that the coordinates of the transformed point
depend only on $\pi^\alpha(x)$, i.e. on the parameters of the
transformations corresponding to the broken generators. The rest
of the transformations (corresponding to the $H$ subgroup) leave
the vacuum unchanged and therefore they are not GB. Thus not all
the isometries  will give rise to zero modes of the brane. When
the $B$ space is homogeneous, the isotropy group does not depend
on the particular point we choose, i.e, $H(Y_0)=H$ and it is
possible to prove that $B$ is diffeomorphic to $K=G(B)/H$, i.e.
the number of GB equals the dimension of $B$. 

According to the previous discussion, we can write the brane
effective action in terms of branon fields using:
\begin{equation}
\partial_{\mu}Y^m(x)=\frac{\partial Y^m}{\partial
\pi^\alpha}\partial_\mu \pi^\alpha=
\frac{1}{kf^2}\xi^m_\alpha(Y_0)\partial_{\mu}\pi^\alpha+\Od(\pi^2)
\end{equation}
and, therefore, introducing the tensor $h_{\alpha\beta}(\pi)$  as
\begin{equation}
h_{\alpha\beta}(\pi)=f^4 \tilde g'_{mn}(Y(\pi))\frac{\partial
Y^m}{\partial\pi^\alpha}\frac{\partial Y^n}{\partial\pi^\beta},
\end{equation}
we have
\begin{equation}
g_{\mu\nu}=\tilde
g_{\mu\nu}-\frac{1}{f^4}h_{\alpha\beta}(\pi)\partial_{\mu}\pi^\alpha
\partial_{\nu}\pi^\beta
\end{equation}

Thus, for small brane excitations in a background metric $\tilde
g_{\mu\nu}$, the effective action becomes
\begin{equation}
S_{eff}[\pi]=S_{eff}^{(0)}[\pi]+ S_{eff}^{(2)}[\pi]+
S_{eff}^{(4)}[\pi]+ ...
\end{equation}
where:
\begin{eqnarray}
S_{eff}^{(0)}[\pi]&=&-f^4 \int_{M_4}d^4x\sqrt{\tilde g}.
\end{eqnarray}
The $\Od(p^2)$ contribution is the non-linear sigma model
corresponding to a symmetry-breaking pattern $G\rightarrow H$ plus
the background scalar curvature term:
\begin{eqnarray}
 S_{eff}^{(2)}[\pi]&=&\frac{1}{2}\int_{M_4}d^4x\sqrt{\tilde g}
h_{\alpha\beta}\partial_{\mu}\pi^\alpha\partial^{\mu}\pi^\beta
+\lambda f^2\int_{M_4}d^4x\sqrt{\tilde g} \tilde R.
\end{eqnarray}

We are assuming that the branons derivative terms are of the same
order as those with metric derivatives. The fourth-order term is
obtained by expanding both the metric determinant and the induced
scalar curvature in branon fields. Up to a surface term (see \cite{nuestro2}), 
we obtain:
\begin{eqnarray}
 S_{eff}^{(4)}[\pi]&=&\frac{-1}{8f^4}\int_{M_4}d^4x\sqrt{\tilde g}h_{\alpha\beta}
 h_{\gamma\delta}(\partial_{\mu}\pi^\alpha\partial^{\mu}\pi^\beta\partial_{\nu}
 \pi^\gamma\partial^{\nu}\pi^\delta-2\partial_{\mu}\pi^\alpha\partial^{\nu}
 \pi^\beta\partial_{\nu}\pi^\gamma\partial^{\mu}\pi^\delta)\nonumber\\
&+&\frac{\lambda}{2f^2} \int_{M_4}d^4x\sqrt{\tilde
g}h_{\alpha\beta}\partial^{\mu}\pi^\alpha\partial^{\nu}\pi^\beta(2\tilde
R_{\mu\nu}-\tilde R\tilde g_{\mu\nu}), \label{ese4}
\end{eqnarray}

Let us
emphasize again that the above effective action is an expansion in
branon fields (or metric) derivatives over $f^2$ and not an
expansion in powers of $\pi$ fields, i.e, it is a low-energy
effective action. 
Notice also that we have assumed that the $G(M_D)$ symmetry
is exact, which implies that the branon fields are massless.
However, in a real situation, such symmetry will be only
approximately realized. In this case, we will expect the branons
to acquire a mass that will measure the breaking of the $G(M_D)$
symmetry. In \cite{nuestro2} we have studied these effects and 
how the symmetry
breaking affects the brane ground state.

\section{Kaluza--Klein gauge bosons and the Higgs mechanism}

As commented in the introduction, the isometries in the $B$ space
are considered as gauge transformation in the Kaluza--Klein
theories \cite{Balo}. In this section we study under which
circumstances the GB associated to the isometry breaking can give
rise to the longitudinal components of the Kaluza--Klein gauge
bosons, as in the standard Higgs mechanism.

 We start with the Hilbert--Einstein action for the gravitational field
 in $D$ dimensions  plus the brane action: $S=S_G+S_B$, i.e.
\begin{equation}
S=\frac{-1}{16\pi G_D}\int_{M_D} d^Dz\sqrt{G} R_D    -f^4
\int_{M_4}d^4x\sqrt{g}
G_{MN}g^{\mu\nu}\partial_{\mu}Y^M\partial_{\nu}Y^N,
\end{equation}
where $z=(x,y)$ are the coordinates defined on $M_D$, $x$ and $y$
being the coordinates defined on $M_4$ and $B$ respectively, and
$R_D$ is the $D$-dimensional scalar curvature. 

In a general non-Abelian case, we consider the usual
ansatz for the metric tensor used in the 
Kaluza--Klein theories:
\begin{eqnarray*}
 G_{MN}&=&
\left(
\begin{array}{cccc}
\tilde g_{\mu\nu}(x)-g'_{mn}(y)B_{\mu}^m(x,y)B_{\nu}^n(x,y)&B_{\mu}^n(x,y)\\
B_{\nu}^m(x,y)&-g'_{mn}(y)
\end{array}\right),
\end{eqnarray*}
where $B_{\mu}^n(x,y)=\xi^n_a(y)A_{\mu}^a(x)$ with $\xi^n_a(y)$ the
Killing vectors
 corresponding to the isometry group $G(B)$ introduced above. Now the
gauge transformations are $y^m \rightarrow y'^m=y^m+\xi^m_a(y)\epsilon^a(x)$.
As is well known in this case, the gravitational action $S_G$ can be written
as
\begin{equation}
S_G=\frac{-1}{16\pi G_N}\int d^4x\sqrt{\tilde g} \tilde R- \frac{\langle
\xi^n_a\xi^m_b g'_{mn}\rangle }{16\pi G_N}
\frac{1}{4} \int_{M_4}d^4x \sqrt{\tilde g} F^a_{\mu\nu}F^{\mu\nu
b},
\end{equation}
where
$F^a_{\mu\nu}=\partial_{\mu}A^a_{\nu}-\partial_{\nu}A^a_{\mu}-
C_{abc}A^b_{\mu} A^c_{\nu}$ and
\begin{equation}
16 \pi G_N=\frac{16 \pi G_D}{\int_{B}d^{D-4} y  \sqrt{g'}}.
\end{equation}
In order to obtain the standard Yang--Mills action, the
normalization of the Killing vectors is given by
\begin{equation}
\langle\xi_a^m(y)\xi_b^n(y)g'_{mn}(y)\rangle=k^2\delta_{ab},  \label{norm}
\end{equation}
where $g'_{mn}(y)$ is the $B$ space-time metric, again $k^2=16\pi G_N$
and the brackets are defined as the $B$ manifold average
\cite{Weinberg}
\begin{equation}
\langle f(y) \rangle= \frac{ \int_{B}d^{D-4} y\sqrt{g'}f(y) }{
\int_{B}d^{D-4}y\sqrt{g'} }.
\end{equation}

The induced metric in this case with non-vanishing gauge fields is 
given by:
\begin{equation}
g_{\mu\nu}=\tilde g_{\mu\nu}-\Delta_{\mu}Y^m\Delta_{\nu}Y^ng'_{mn}
\end{equation}
where
the covariant derivative is defined as
\begin{equation}
\Delta_{\mu}Y^m=\partial_{\mu}Y^m-\xi^m_aA_{\mu}^a,
\end{equation}
which can be written as
\begin{equation}
\Delta_{\mu}Y^m=\frac{\partial Y^m}{\partial \pi^\alpha}
\partial_{\mu}\pi^\alpha-\xi^m_a A_{\mu}^a=
\frac{1}{kf^2}\xi^m_\alpha(Y_0) (\partial_\mu \pi^\alpha-kf^2
A_\mu^\alpha)-\xi^m_i(Y_0) A_\mu^i+\Od(\pi^2).
\end{equation}
Since the $i$ indices correspond to the generators of the isotropy
group $H(Y_0)$, the Killing fields vanish at $Y_0$, i.e.
$\xi_i^m(Y_0)=0$ and the last term vanishes.

Therefore the brane action $S_B$ is
\begin{equation}
S_B=-f^4 \int_{M_4}d^4x\sqrt{\tilde g}
+\frac{1}{2}\int_{M_4}d^4x\sqrt{\tilde g} \tilde
g^{\mu\nu}h_{\alpha\beta}D_{\mu}\pi^\alpha
D_{\nu}\pi^\beta+\Od(\pi^4).
\end{equation}
where $D_\mu\pi^\alpha=\partial_\mu \pi^\alpha-kf^2 A_\mu^\alpha$.
Thus the gauge boson mass matrix is
\begin{equation}
M_{\alpha\beta}^2=k^2f^4h_{\alpha\beta}(0).
\end{equation}
Remember that $Y^m(x)=Y^m_0$ corresponds to $\pi^a=0$. As
commented on before, not all the the gauge bosons will acquire a mass
by this mechanism. Only those associated to the broken $X_\alpha$
generators will, their number being determined by the dimension of
the $K=G/H$ space. In any case, two important comments are in
order. First,  the gauge boson
masses are quite small whenever $f \ll M_P$. On the other hand it
should be remembered that in the standard KK picture, having gauge
couplings $g$ small enough to have a sensible weak coupling limit
(say $g<1$) requires having extra dimensions  of a typical size
of the order of the Planck length, since $g^2$ is of the order of
$k^2/R^2$ (see for instance \cite{Balo}). Thus, for the
interesting case of large extra dimensions and $f \ll M_D$,
graviphotons can be considered massless and decoupled from the
rest of the low-energy particles. In this case we can safely
assume that the Higgs mechanism has not taken place and the GB can
be considered as the only relevant low-energy new degrees of
freedom.

\section{Couplings to the Standard Model fields}
As we have shown in the previous sections, the induced metric on
the brane depends on both the four-dimensional bulk metric
components $\tilde g_{\mu\nu}$ and the Goldstone bosons
$\pi^\alpha$. In the following, we will assume that the physical
space-time metric is $\tilde g_{\mu\nu}$,  whereas the contribution
of the GB to the brane metric can be detected only through their
couplings to the Standard Model fields. In order to obtain such
couplings, we use the Sundrum effective action for the SM
fields \cite{Sundrum}, which is basically the SM action defined on
a curved space-time $M_4$ whose metric is the induced metric
$g_{\mu\nu}$. Let us give the results up to
$\Od(p^2)$ for the different kinds of
fields which have been obtained in \cite{Dobado}:

\subsubsection*{Scalars}
For a scalar field we have:
\begin{eqnarray}
S_{\Phi} &=&\frac{1}{2}\int_{M_4}d^4x\sqrt{g}g^{\mu\nu}
\partial_{\mu}\Phi\partial_{\nu}\Phi=\frac{1}{2}\int_{M_4}d^4x
\sqrt{ \tilde g}\tilde
g^{\mu\nu}
\partial_{\mu}\Phi\partial_{\nu}\Phi    \nonumber \\
&+& \frac{1}{2f^4}\int_{M_4}d^4x\sqrt{ \tilde
g}h_{\alpha\beta}(\pi)
\partial_{\mu}\Phi\partial_{\nu}\Phi\partial^{\mu}\pi^\alpha
\partial^{\nu}\pi^\beta
 \nonumber \\
&-& \frac{1}{4f^4}\int_{M_4}d^4x\sqrt{ \tilde g} \tilde
g^{\mu\nu}\partial_{\mu}\Phi\partial_{\nu}\Phi \tilde g^{\rho\sigma}
h_{\alpha\beta}(\pi)
\partial_{\rho}\pi^\alpha \partial_{\sigma}\pi^\beta +\Od(p^4).
\end{eqnarray}

\subsubsection*{Fermions}

The introduction of fermions on the brane is more complicated and
we refer the reader to the above reference for details. The results
are:
\begin{eqnarray}
S&=&\int d^4 x \sqrt{\tilde g} i\bar \psi \tilde{\Dbar} \psi -
\frac{i}{2f^4}\int d^4 x \sqrt{\tilde g}h_{\alpha\beta}(\pi)\tilde
g^{\mu\nu}\partial_\mu \pi^\alpha\partial_\nu\pi^\beta\bar \psi
\tilde{\Dbar} \psi\nonumber \\
&+&\frac{i}{2f^4}\int d^4 x \sqrt{\tilde g} \bar \psi
h_{\alpha\beta}(\pi)\tilde g^{\mu\nu}\partial_\mu
\pi^\alpha\pabar\pi^\beta\tilde D_\nu\psi
\nonumber
\\&-&\frac{i}{4f^4}\int d^4 x \sqrt{\tilde g} \bar \psi
h_{\alpha\beta}(\pi)\tilde g^{\mu\nu}(\pabar(\partial_\mu
\pi^\alpha\partial_\nu\pi^\beta) -\partial_\mu(\pabar
\pi^\alpha\partial_\nu\pi^\beta))\psi.
\end{eqnarray}
where $\tilde D_\mu$ is the fermionic covariant derivative corresponding
to the
background metric $\tilde g_{\mu\nu}$. 
In particular, this way of introducing the couplings of Goldstone
bosons to fermions allows us to consider chiral fermions in a
straightforward way.

\subsubsection*{Gauge bosons}

For the Yang--Mills action on the brane we can follow similar
steps, and we get:
\begin{eqnarray}
S_{YM}&=&\frac{\tr}{2g^2}\int d^4x \sqrt{\tilde g} \tilde
g^{\mu\rho}\tilde g^{\nu\sigma} G_{\mu\nu} G_{\rho\sigma}\nonumber
\nonumber\\&-& \frac{\tr}{4g^2f^4}\int d^4x \sqrt{\tilde g} \tilde
g^{\mu\nu}h_{\alpha\beta}(\pi)(\partial_\mu \pi^\alpha\partial_\nu
\pi^\beta)\tilde g^{\mu\rho}\tilde g^{\nu\sigma} G_{\mu\nu}
G_{\rho\sigma}\nonumber
\\ &+&\frac{\tr}{g^2f^4}
\int d^4x \sqrt{\tilde g}h_{\alpha\beta}(\pi)(\partial_\lambda
\pi^\alpha\partial_\kappa \pi^\beta)
\tilde g^{\mu\lambda}\tilde g^{\rho\kappa}\tilde g^{\nu\sigma}
G_{\mu\nu} G_{\rho\sigma}
\end{eqnarray}
From the above discussion we see that the GB always interact by pairs
with
the SM matter. In addition, due to their geometric origin,
those interactions are very similar to the gravitational interactions
since the $\pi$ fields couple to all the matter fields with
the same strength, which is suppressed by a
factor
$f^4$. This is quite interesting since it could explain why they
have
 not
been observed so far, provided they exist at all. However, moderate
values
 of the brane tension around the TeV scale could make their
 production possible in the next generation of colliders.

\section{Brane-skyrmions}

The branon fields introduced in the previous sections describe
small oscillations of the brane around its ground state. Thus
there is some similitude with the well known chiral lagrangian
approach in which a non-linear sigma model (NLSM) is used to
describe the low-energy pion dynamics. Apart from pions, the NLSM
can also be used to describe other non-trivial states in the
hadron spectrum such as baryons. For that purpose, the non-trivial
topological structure of the coset space $K$ plays a fundamental
role.  In fact, baryons can be identified with certain
topologically non-trivial maps between the (compactified) space
$S^3$ and the coset manifold $K$ known as Skyrmions.

Let us then consider static branon field configurations with
finite energy, that accordingly vanish at the spatial infinity.
Thus, we can compactify the spatial dimensions to $S^3$ and the
static configurations will be mappings: $\pi^\alpha:
S^3\rightarrow K$. Therefore, these mappings can be classified
according to the third homotopy group of $K$, i.e. $\pi_3(K)$. As
a consequence, mappings belonging to different non-trivial
homotopy classes cannot be deformed in a continuous fashion from
one to the other. This implies that such configurations cannot
evolve in time classically into the trivial vacuum $\pi=0$ and
therefore they are stable states. For the sake of simplicity we
will consider the case in which we have $N=3$ extra dimensions
with $B=S^3$. In this case, since $B$ is an homogeneous space, we
have $K\sim B=S^3 \sim SU(2)$, i.e the coset manifold is also a
3-sphere. Thus, we will have $\pi_3(S^3)=\bf{Z}$ and the mappings
can be classified by an integer number, usually referred to as the
winding number $n_W$. We will also assume in the following, that
the background metric is flat, i.e, $\tilde g_{\mu\nu}=
\eta_{\mu\nu}$.

For static configurations the  brane-skyrmion mass can be
obtained directly from the effective lagrangian as:
\begin{eqnarray}
M[\pi]=-\int d^3 x {\cal L}_{eff}.
\end{eqnarray}
In general this expression will be divergent because of the volume
contribution coming from the lowest order term
$S^{(0)}_{eff}[\pi]$ which reflects the fact that the brane has
infinite extension with  finite tension. Therefore, in order to
obtain a finite skyrmion mass, we will substract the vacuum energy
$M[0]$, i.e.:
\begin{equation}
M_S[\pi]=M[\pi]-M[0]=f^4 \int_{M_3}d^3x\sqrt{g}-\lambda f^2
\int_{M_3}d^3x\sqrt{g}R-M[0]. \label{mass}
\end{equation}
In other words we are defining the mass of the brane-skyrmion as
the mass of the brane with the topological defect minus the mass
of the brane in its ground state with no topological defect.

In order to simplify the calculations we will introduce spherical
coordinates on both spaces, $M_4$ and $K$. In $M_4$ we denote the
coordinates $\{t,r,\theta,\varphi\}$ with $\phi \in [0,2\pi)$,
$\theta \in[0,\pi]$ and $r \in [0,\infty)$. 
On  the coset  manifold $K$, the spherical coordinates are denoted
$\{\chi_K,\theta_K,\phi_K\}$ with $\phi_K \in [0,2\pi)$, $\theta_K
\in [0,\pi]$ and $\chi_K \in [0,\pi]$. These coordinate are
related to the physical branon fields (local normal geodesic
coordinates on $K$) by:
\begin{eqnarray}
\pi_1&=&v \sin\chi_K\sin\theta_K\cos\phi_K,\nonumber\\
\pi_2&=&v \sin\chi_K\sin\theta_K\sin\phi_K,\label{esfk}\\ \pi_3&=&v
\sin\chi_K\cos\theta_K.\nonumber
\end{eqnarray}

The coset metric $h_{\alpha\beta}$ is the usual $S^3$ metric
in spherical coordinates.
In these coordinates, the brane-skyrmion with winding number
$n_W$ is given by the non-trivial mapping
$\pi^\alpha:S^3\longrightarrow S^3$ defined from:
\begin{eqnarray}
\phi_K&=&\phi,\nonumber
\\ \theta_K&=&\theta,\nonumber\\
\chi_K&=&F(r),\label{skyan}
\end{eqnarray}
with boundary conditions $F(0)=n_W\pi$ and $F(\infty)=0$. This
map is usually referred to as the hedgehog ansatz (see Fig. 2).

\begin{figure}
\vspace*{0cm}
\centerline{\mbox{\epsfysize=5.5 cm\epsfxsize=5.5
cm\epsfbox{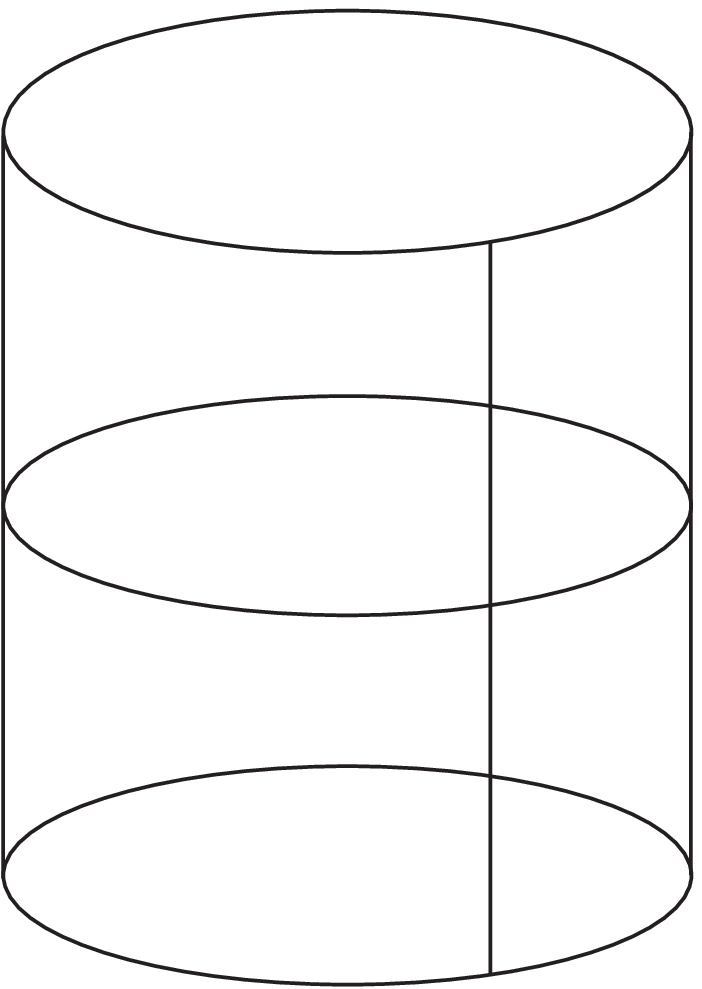}}\hspace*{2 cm}\mbox{\epsfysize=5.5
cm\epsfxsize=5.5 cm\epsfbox{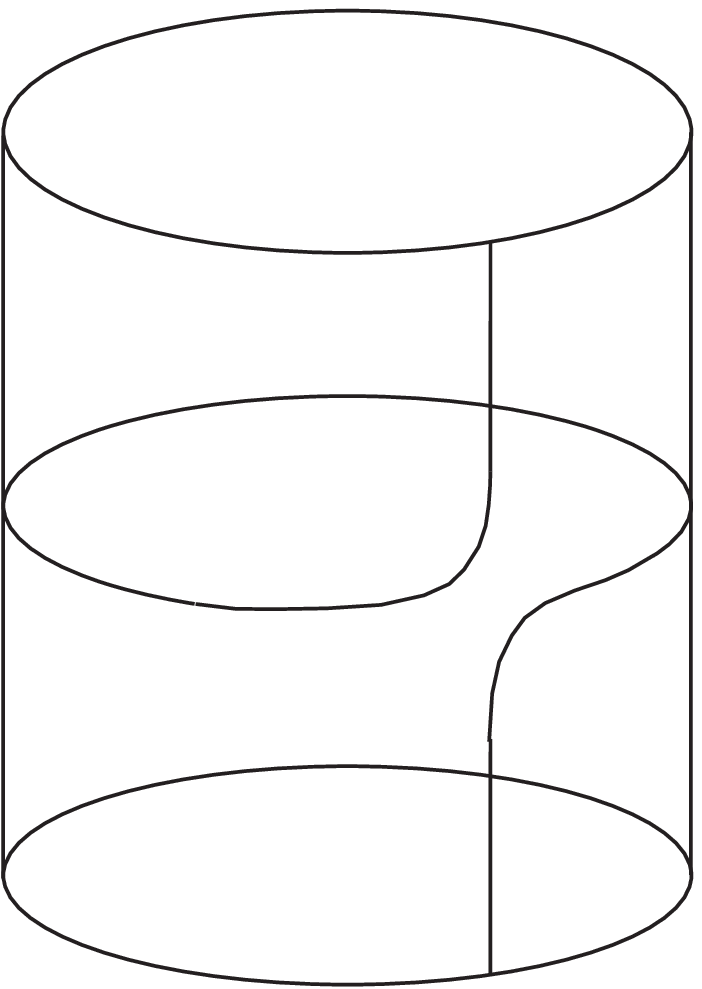}}}
%
 \caption{\footnotesize{Brane configurations with $n_W=1$
in $M_3=M_2\times S^1$. On the right we plot  a non-zero size
skyrmion. On the left a  zero-size 1-brane-skyrmion  is shown. It
has the same mass and  shape as a wrapped soliton and a
topologically trivial (world) brane. However the topology is not
the same.}}
\end{figure}

In order to calculate the brane-skyrmion mass from (\ref{mass}),
we need explicit expressions for the induced metric determinant
and the scalar curvature in terms of the profile function $F(r)$, so that 
 we can write $M_S$ as a functional of $F(r)$. The actual mass
of the brane-skyrmion with winding number $n_W$ will be obtained
by minimizing the functional $M_S[F]$ (\ref{mass}) in the space
of functions $F(r)$ with the appropriate boundary conditions. This problem is
in general rather complicated, but we can obtain an upper bound to
the mass by using a family of functions parametrized by a single
parameter, and minimizing with respect to that parameter. In
particular, it is very useful to work with the Atiyah-Manton
ansatz \cite{Atiyah}:
\begin{eqnarray}
 F(r)=n_W \pi\left(1-\frac{1}{\sqrt{1+L^2/r^2}}\right).
\label{AM}
\end{eqnarray}
By minimizing the brane-skyrmion mass with
respect to $L$ we will get: $M_S\equiv min_L\; M_S(L)\equiv M_S(L_m)$
for different values of the parameter $\lambda$. Thus:
\begin{itemize}
\item For
$\lambda=0$ and using the ansatz above for $n_W=1$ we find that
$M_S(L)$ is minimized for $L_m=0$. The corresponding mass is given
by
\begin{equation}
M_S=2\pi^2f^4R_B^3,
\end{equation}
i.e. the brane-skyrmion describes a
pointlike particle with a finite mass given by the volume of the
extra dimensions times the brane tension $f^4$. It can be shown 
that this result is general, i.e. it is valid
for any parametrization of the $F(r)$ function and not only for
the Atiyah-Manton one.

\item For $\lambda>0$, we find that $M_S(L)$ is minimized for some $L_m>0$
(non-zero size brane-skyrmion). In this case, assuming that
 the contribution
from the curvature term never becomes negative, it is
possible to obtain the following lower bound on the mass
$M_S>M_S(\lambda=0)=2\pi^2f^4R_B^3$. An upper bound can be
obtained evaluating $M_S$ in the limit $L=0$,
simply assuming that this limit is well defined so that
we can commute the limit with the integration. Thus
\begin{equation}
M_S<M_S(L=0)=2\pi^2f^4R_B^3\left(1+6\frac{\lambda}{R_B^2f^2}\right).
\end{equation}
These results are general for any monotonic parametrization and, in particular,
we have checked numerically that they hold for the Atiyah-Manton case.

\item Finally, for $\lambda<0$, we find again that the minimum $M_S$
corresponds to zero size brane-Skyrmion and the corresponding mass
is
\begin{equation}
M_S=2\pi^2f^4R_B^3\left(1 +6\frac{\lambda}{R_B^2f^2}\right).
\end{equation}
When $\lambda<-R_B^2f^2/6$ the brane-skyrmion mass becomes
negative. In this case, using non-monotonic parametrizations, we
have obtained that the mass  is actually not bounded from below,
since the curvature term can be made arbitrarily large. As a
consequence only in this last case, the brane-skyrmion becomes
unstable. These results are summarized in Table.1
\end{itemize}
\begin{table}[h]
\begin{center}
\begin{tabular}{|c|c|c|}
\hline
 & & \\
 $\lambda$& Size &   Mass \\
 & & \\
\hline $\lambda=0$& $L_m=0$ & $M_S=2\pi^2f^4R_B^3$\\ \hline
$\lambda>0$ &$L_m>0$ & $2\pi^2f^4R_B^3<M_S<2\pi^2f^4R_B^3(1
+6\frac{\lambda}{R_B^2f^2})$ \\ \hline $-R_B^2f^2/6<\lambda<0$ &
$L_m=0$ & $M_S=2\pi^2f^4R_B^3(1 +6\frac{\lambda}{R_B^2f^2})$ \\
\hline $\lambda<-R_B^2f^2/6$ & $L_m=0$ & $M_S\rightarrow
-\infty$\\ \hline
\end{tabular}
\caption{\footnotesize{Values of the size $L_m$ and mass $M_S$ for
the brane-skyrmion with $n_W=1$ for different values of the
$\lambda$ parameter.}}
\end{center}\label{masas}
\end{table}
In Fig.3 we show the brane-skyrmion mass as a function of
$\lambda$ and $L$. We see that the minimum is displaced from $L=0$
when $\lambda>0$.
\begin{figure}[h]
\vspace*{0cm}
\centerline{\mbox{\epsfysize=10 cm\epsfxsize=10
cm\epsfbox{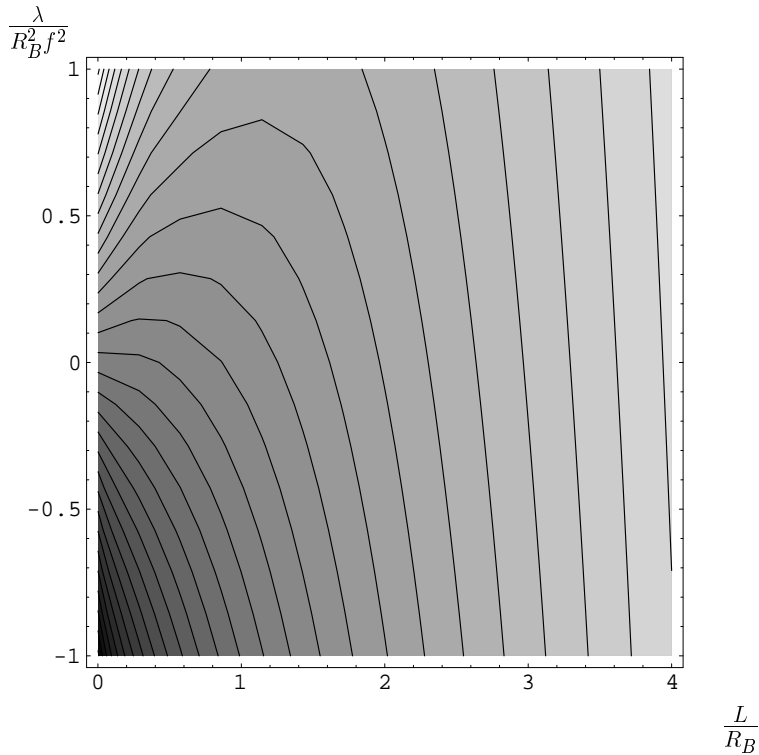}}}
%
 \caption{\footnotesize{Contour plot of the brane-skyrmion mass as a
function of
 $\lambda/(R_B^2f^2)$ and $L/R_B$.}}
\end{figure}

For more precise numerical results see \cite{nuestro2}. 
An interesting conclusion that can be obtained from those results
is that the interaction between two classical brane-skyrmions with
topological number $n_W=1$ is repulsive when their sizes are non
zero, whereas they do not interact if their sizes are exactly
zero. 

We can also take into account the effects of the possible branon mass
on the Skyrmion properties. As shown in \cite{nuestro2}, the
presence of the branon masses does not affect the brane-skyrmion
stability, the only
difference with the massless case being that the brane-skyrmion mass
increases.

Finally this analysis can be extended to higher-dimensional Skyrmions.
For the simple case where $M_D=M_{N+1}\times S^N$ similar results hold,
except for $N=1$ where brane-Skyrmions are unstable, and $N=2$ where
the contribution of the curvature term vanishes.

\section{Interaction Lagrangian and fermionic quantization}

In this section we will study the interaction between the
brane-skyrmions and the branons. For simplicity we will consider
only the case $M_7=M_4\times S^3$ with $\tilde
g_{\mu\nu}=\eta_{\mu\nu}$ so that $K \simeq SU(2)$. Then we can
follow the well known steps for quantization of the standard
chiral dynamics Skyrmion \cite{Witten1}. It is possible to split
the isometry group $G$ as $G=SU(2)_L \times SU(2)_R$ and $H$
corresponds to the isospin group $SU(2)_{L+R}$. The
parametrization of the coset is usually done in terms of a $SU(2)$
matrix $U(x)$ and the Skyrmion   is usually written as
\begin{eqnarray}
U(x)&=&\exp (iF(r)\hat x^a \tau^a)=\cos F(r) + i\tau^a \hat x^a
\sin F(r)\nonumber \\ &=&\pm\sqrt{1 -\frac{\pi^2}{v^2}}+i\tau^a
\hat x^a\sin F(r),
\end{eqnarray}
where $\tau^a$ are the $SU(2)$ generators. From this expression we
can identify the Goldstone bosons fields $\pi^\alpha=v \sin F(r)
\hat x^\alpha$. The quantization of the isorotations of the
Skyrmion  solution (which correspond to rotations in the
compactified space $B=S^3$ in our case) requires the well-known
relation $J=I$ where $J$ and $I$ are the spin and the isospin
indices. In principle the allowed values of $J$ are
$J=0,1/2,1,3/2,...$. As explained by Witten \cite{Witten2},
fermionic quantization is possible because of the
Wess-Zumino-Witten (WZW) term $k \Gamma $ with $k$ integer, that
can be added to the Goldstone boson effective action. For $k$ even
the Skyrmion is a boson, but for $k$ odd it is a fermion. For the
$SU(2)$ case, the functional $\Gamma$ has no dynamics and becomes
a topological invariant related with the homotopy group
$\pi_4(SU(2))= \bf{Z}_2$. Note that for a suitable
compactification of the space-time this is the relevant group for
the Goldstone boson map. A map belonging to the non-trivial class
could for example describe the creation of a Skyrmion-antiskyrmion
pair followed by a $2\pi$ rotation of the Skyrmion and finally a
Skyrmion-antiskyrmion annihilation. In the fermionic case this
field configuration must be weighted with a $-1$ in the Feynman
path integral. For an adiabatic $2\pi $ rotation of the Skyrmion
around some axis, the WZW term contributes $k \pi$ to the action
and $(-)^k$ to the amplitude which can be understood as an $exp(i
2 \pi J)$ factor. Therefore both possibilities, bosonic and
fermionic quantization of the Skyrmion, are open. In principle
this result can also be extended to the more general case of $S^N$
brane-skyrmions considered in the last section where
$M_D=M_{N+1}\times S^N$ with $D=2 N +1$ since
$\pi_{N+1}(S^N)=\bf{Z}_2$ for $N \geq 3$.

In order to study the low-energy interactions of the
brane-skyrmions with the branons we have to obtain the appropriate
effective lagrangian. This lagrangian must be $G(B)$ symmetric and
the brane-skyrmion is described in it by a complex field because
of the topological charge. Thus for example this field will be a
complex scalar $\Phi$ for $J=0$ or a Dirac spinor $\Psi$ for
$J=1/2$. For the scalar case the invariant  lagrangian with the
lowest number of derivatives can be written as:
\begin{eqnarray}
{\cal L}_{s}=\alpha \Phi^* \Phi h_{\alpha\beta}(\pi)\partial_\mu
\pi^\alpha
\partial^\mu \pi^\beta.
\end{eqnarray}
The coupling $\alpha$ can be obtained from the large distance
behaviour of the branon field in the brane-skyrmion configuration, 
i.e., $F(r)\simeq B/r^2$.
In particular for the  Atiyah-Manton ansatz with $n_W=1$, we
get $B=L^2 \pi/2$.
 By using the
lagrangian ${\cal L}_{s}$ it is also possible to obtain the branon
field produced by the brane-skyrmion field $\Phi$ and by
comparison with the above results we arrive at \cite{Clements}
\begin{equation}
\alpha=-\frac{8}{3}\pi^2v^2B^2=-\frac{2}{3}v^2\pi^4 L_m^4.
\end{equation}
From this lagrangian it is possible, for example, to compute the
cross sections for producing a brane-skyrmion-antibrane-skyrmion
pair from two branons.

The fermionic case can be studied in a similar way \cite{Clements,Witten1}
although a consistent analysis would be more involved, since it
requires the quantization of the rotational modes. 

\section{Wrapped states}

In this section we introduce another kind of states which
can appear as topological excitations of the branes. These states
correspond to brane configurations  wrapped around the
compactified spaces $B$ which typically will be assumed to be
$S^N$ for $M_D=M_{N+1}\times B$. A given wrapped states is located
at some well defined point of the space $M_N$. The possibility of
having this wrapped states is related to the homotopy group
$\pi_N(B)=\bf{Z}$ which is obviously the case for $B=S^N$ but also
for other spaces. In principle, wrapped states can be present even
when there is no world brane, i.e., when we do not have a brane
extended along the big space $M_N$. However, one of the most
interesting cases occurs when the wrapped states are located at
one point of the world brane and then can be understood as
world-brane excitations. Note that as far as the relevant homotopy
group is again $\bf{Z}$ we have also antiwrapped states which
correspond to negative winding numbers. Thus a world brane can get
excited by creating a wrapped-antiwrapped state at some given
point. In Fig. 4 we show a single wrapped state at rest (left) and
branon-excited. On the left of Fig.2 we show a wrapped state
(circle) located at one point of the world brane (straight line)
for the case $N=1$.

To study in more detail the main properties of these wrapped
states we concentrate now on a four-dimensional space-time $M_4$
embedded in a $7$-dimensional bulk space that we are assuming to
be $M_7=M_4\times B$ with $M_4={\bf R}\times M_3$ and $B=S^3$.
Now, unlike the brane-skyrmion case, the finite energy requirement
do not lead to any compactification of the world space $M_3$
because the brane is going to be wrapped around $B$ which is
compact. However, for technical reasons it is still useful to
compactify $M_3$ to $S^3$ by adding the spatial infinite point.
The wrapped brane produces the spontaneous breaking of the $M_7$
isometry group, which we assume to be $G(M_7)=G({\bf R}\times
M_3\times S^3)=G({\bf R}\times S^3)\times G(M_3)$ to the $G({\bf
R}\times S^3)\times H'$ group where $H'$ is the isotropy group of
$M_3$ which is assumed to be homogeneous. Then the coset space is
defined by $K'=G(M_3)/H'$. In the simplest case $M_3=S^3$ we have
$K'=SO(4)/SO(3)=S^3$ and then $K'\sim B\sim S^3$. Therefore the
low-energy brane excitations can be parameterized as

\begin{figure}
\vspace*{0cm}
\centerline{\mbox{\epsfysize=5.5 cm\epsfxsize=5.5
cm\epsfbox{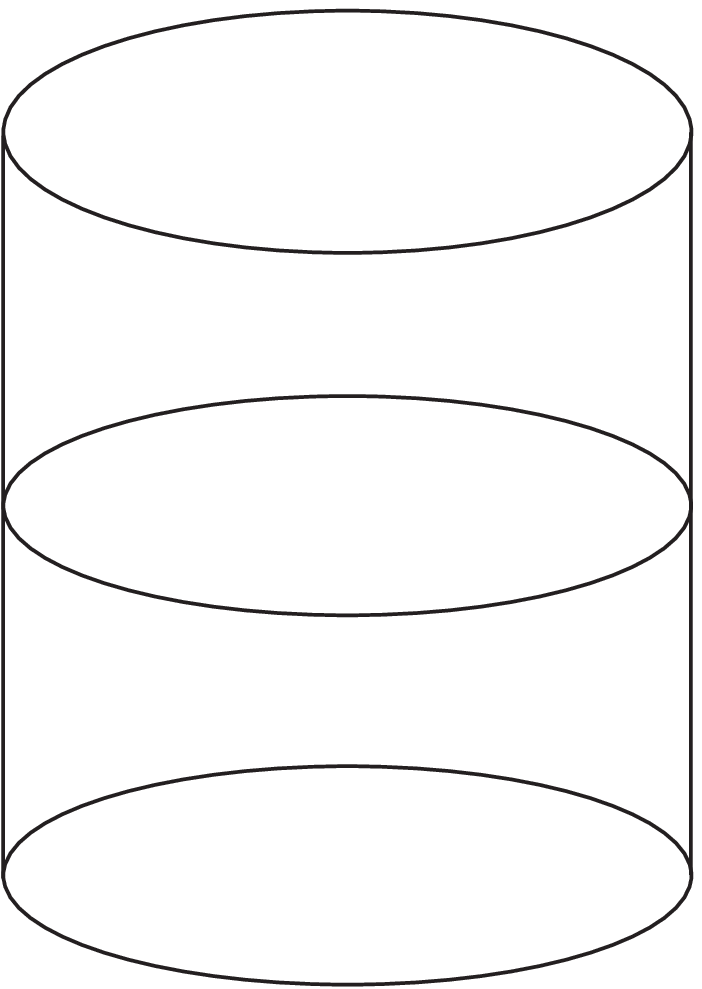}}\hspace*{2 cm}\mbox{\epsfysize=5.5
cm\epsfxsize=5.5 cm\epsfbox{cilwonp1.eps}}}
%
 \caption{ \footnotesize{Wrapped brane with topology number 1 in a
1-dimensional compact extra-space in addition to 1-dimensional
space. The ground state of the brane is represented in the left
graphic, and a exited state is represented in the right graphic.}}
\end{figure}

\begin{eqnarray}
\Pi:B&\longrightarrow&K'\nonumber\\ y &\longrightarrow&\pi(y),
\end{eqnarray}
where  $y$ are coordinates on $B$ and $\pi$ are coordinates on
$K'$. On the other hand, as far as the quotient space $K'$ and
$M_3$ are both topologically equivalent to $S^3$, it is possible
to describe the wrapped brane by giving its position on the $M_3$
space $X^i$ as a function of $y^m$, i.e. $X^i=X^i(y^m)$. In
particular it is possible to choose the coordinates so that
\begin{equation}
X^i(y)=\frac{1}{f^2}\delta^i_\alpha \pi^\alpha(y)+...
\end{equation}
locally. In the following we will use $X^i$ instead of
$\pi^{\alpha}$ to label the wrapped-brane points in terms of the
brane parameters $y$ . Let us now rearrange the coordinates for
convenience in order to have $Y^M=(t,y^m,X^i(y))$ where $t$ is the
temporal coordinate $t=x^0$. The bulk metric is then:
\begin{eqnarray}
 G_{MN}&=&
\left(
\begin{array}{cccc}\tilde g_{00}&&0\\
&-\tilde g'_{mn}(y)&\\ 0&&-\tilde g_{ij}(x)
\end{array}\right)=\left(
\begin{array}{cccc}\tilde g'_{rs}(y)&0\\ 0&-\tilde g_{ij}(x)
\end{array}\right),
\end{eqnarray}
where $\tilde g'_{rs}$ is the background metric on the space-time
manifold ${\bf R}\times B$, i.e., $r,s=0,1,2,3$. The induced
metric on this manifold can be evaluated in a way similar to the
brane-skyrmion case. Thus in the ground state, the induced metric
on the wrapped brane is given by the four-dimensional components
of the bulk space metric, i.e. $g'_{rs}=\tilde g'_{rs}=G_{rs}$.
When branons are present, the induced metric is given by
\begin{equation}
g'_{rs}=\tilde g'_{rs}-\partial_{r}X^i\partial_{s}X^j\tilde g_{ij}
\end{equation}
On the other hand, the action including the scalar curvature term is
given by:
\begin{equation}
S_B=-f^4 \int_{{\bf R}\times B}dtd^3y\sqrt{g'}+\lambda f^2
\int_{{\bf R}\times B}dtd^3y\sqrt{g'}R',\label{wra}
\end{equation}
where $R'$ is the induced curvature on the wrapped brane
and the volume term is now finite for fixed time. For
small excitations, the effective action becomes
\begin{equation}
S_{eff}[X]=S_{eff}^{(0)}[X]+ S_{eff}^{(2)}[X]+ ...
\end{equation}
Where the effective action for the branons up to $\Od(p^2)$ is
nothing but the non-linear sigma model corresponding to a
symmetry-breaking pattern $G(M_3)\rightarrow H'$:
\begin{eqnarray}
S_{eff}^{(0)}[X]&=&-f^4  \int_{{\bf R}\times B}dtd^3y\sqrt{\tilde
g'},
\end{eqnarray}
\begin{eqnarray}
 S_{eff}^{(2)}[X]&=&\frac{f^4}{2} \int_{{\bf R}\times B}dtd^3y
\sqrt{\tilde g'}\tilde
 g_{ij}\tilde
 g'^{rs}\partial_{r}X^i\partial_{s}X^j
+\lambda f^2 \int_{{\bf R}\times B}dtd^3y\sqrt{\tilde g'} \tilde
R'.\hspace{1 cm}
\end{eqnarray}
Where $\tilde R'$ is the background curvature on the wrapped
brane, without excitations. Notice that $i, j,...$  are $M_3$
indices, whereas $r, s,...$ are indices on the ${\bf R}\times K'$
manifold. This effective action is again an expansion in powers of
$p \sim \partial_r X\sim\partial_r g'/f$, i.e. it is a low-energy
expansion. 

For static configurations the mass of the wrapped state, to the lowest
order, is given from (\ref{wra}) again by
\begin{equation}
M_W=f^4 \int_{B}d^3y\sqrt{g'}.
\end{equation}
The minimum is found for $X^i=0$:
\begin{eqnarray}
M_W=2\pi^2f^4R_B^3,
\end{eqnarray}
which is proportional to the $B=S^3$ volume as expected. In this
case we have the brane wrapped around $B$ with the minimal
possible brane volume. For small enough $\lambda$,
adding the curvature term does not change the picture very much:
\begin{equation}
M_W=f^4 \int_{B}d^3y\sqrt{g'}-\lambda f^2 \int_{B}d^3y\sqrt{g'}R'.
\end{equation}
Because of the scalar curvature on a 3-sphere is $\tilde
R'=-6/R_B^2$, we find
\begin{eqnarray}
M_W=2\pi^2f^4R_B^3\left(1+6\frac{\lambda}{R_B^2f^2}\right).
\end{eqnarray}
Then the brane is still wrapped minimizing its volume but we have
a new contribution to the mass coming from the brane curvature
which coincides with the $B$ curvature. This result obviously
applies to branes wrapped once around $B$. The generalization to
cases where the brane is wrapped $n_W\in \bf{Z}$ times is
straightforward  resulting just in a factor of $\vert n_W \vert$
in the above equation.

It is very interesting to realize that the obtained value for
the wrapped-state mass is exactly the same previously given for
the brane-skyrmion mass in Table.1, as the upper bound for
positive $\lambda$ and the exact value for negative $\lambda$,
provided $\lambda > -R_B^2 f^2/6$. The fact that our brane
action is  defined in an entirely geometrical way, makes possible
to give a beautiful explanation of this fact. 
On the right of  Fig.2 we have represented a brane-skyrmion
corresponding to a positive value of $\lambda$. According to our
previous discussion the brane-skyrmion has a non-zero size. This
makes possible to pass through the brane from one side to the
other, showing the  topological defect as some kind of hole in the
brane. 

Despite the similarity between both
configurations, they are not the same because  their topology is
different. The brane-skyrmion  is extended on  both the
compactified $M_3$ space and the extra-dimensional space $B$, but
the wrapped states do only around the extra dimension space $B$.
Another way to understand why they are different is to
realize that the brane-skyrmion is made of a single piece unlike
the wrapped configuration which has two different pieces (the
wrapped brane and the world brane). Thus they cannot be connected
by a classical process, although quantum tunneling could produce
in principle transitions between one to the other.

\section{Summary and conclusions}
In this work we have studied the effective action describing the
low-energy dynamics of the GB, which appear when the
higher-dimensional
space-time manifold isometry group is spontaneously
broken by the presence of a three-brane Universe. From the
$3+1$-dimensional
point of view, those GB can be considered as some kind
of new scalar fields whose dynamics is given by the non-linear
sigma model lagrangian corresponding to the coset manifold
$K=G/H$. Eventually, the GB can also get some mass terms due to
small deviations from the simple ideal exact isometry pattern.

This spontaneous symmetry breaking gives rise, through the Higgs
mechanism, to a mass matrix for the KK graviphotons associated to
the isometries of the compactified space $B$. However, for the
interesting case of large extra dimensions and $f \ll M_D$, the
graviphotons decouple from the low-energy theory and their masses
become very small. We can thus consider the GB as the only
relevant degrees of freedom on the brane in the low-energy regime.

In order to make further studies of the possible phenomenological
implications of those GB brane excitations, we have considered
their corresponding couplings with the SM particles including
scalars, fermions (chiral and non-chiral) and gauge bosons.

Under suitable assumptions about the third homotopy group of the
space $K$, this effective action gives rise to a new kind of states
corresponding to topological defects of the brane
(brane-skyrmions) which are stable whenever the curvature
parameter $\lambda$ is not too negative. The mass and the size of
the brane-skyrmions can be computed in terms of the brane tension
scale ($f$), $\lambda$ and the size of the space $B$ ($R_B$). The
brane-skyrmions can be understood as some kind of holes in the
brane that make possible to pass through them along the $B$ space.
This is because in the core of the topological defect the symmetry
is restablished. In the case considered here the broken symmetry
is basically the translational symmetry along the
extra-dimensions. Thus the core of the brane-skyrmion plays the
role of a window through the brane, which is a nice geometrical
interpretation of this object. For $\lambda =0$ or negative the
brane-skyrmion collapses to zero size
and that window is closed.

Brane-skyrmions can in principle be quantized as bosons or
fermions by adding a Wess-Zumino-Witten-like term to the branon
effective action. This is a very interesting possibility since it
provides a completely new way of introducing fermions on the
brane. The low-energy effective lagrangian describing the interactions
between branons and brane-skyrmions can also be obtained in a systematic
way. This open the door for the study of the possible
phenomenology of these states at the Large Hadron Collider (LHC)
currently under construction at CERN.

We have studied another different set of states corresponding
to a brane wrapped on the extra-dimension space $B$ (wrapped
states) and we have derived their connection with the
brane-skyrmion states.

Finally, these results can be extended also to  higher
dimensions where similar results hold. This fact 
could have some relevance in the context of pure M-theory where
solitonic 5-branes are present which could wrap around
5-dimensional spheres.

We understand that the brane-skyrmions and wrapped states studied
in this paper are quite interesting objects (both from the
theoretical and perhaps from a more phenomenological point of
view) and thus we think that they deserve future research.

\vspace{.5cm}
 {\bf Acknowledgements:} This work
has been partially supported by the Ministerio de Educaci\'on y
Ciencia (Spain) (CICYT AEN 97-1693 and PB98-0782).    \\


\thebibliography{references}
\bibitem{Hamed1} N. Arkani-Hamed, S. Dimopoulos and G. Dvali,
{\it Phys. Lett.} {\bf B429}, 263 (1998)
\bibitem{Sundrum}R. Sundrum, {\it Phys. Rev.} {\bf D59}, 085009 (1999)
\bibitem{KK} T. Kaluza. Sitzungsberichte of the
Prussian Acad. of Sci. 966 (1921)\\ O. Klein, {\it Z. Phys.} {\bf
37}, 895 (1926)
\bibitem{Balo}D. Bailin and A. Love, {\it Rep. Prog. Phys. } {\bf
    50}, 1087
\bibitem{Giudice} G. Giudice, R. Rattazzi and J.D. Wells, {\it
    Nucl. Phys.}
{\bf  B544}, 3 (1999)\\ E.A. Mirabelli, M. Perelstein and M. E.
Peskin, {\it Phys. Rev. Lett.} {\bf 82}, 2236 (1999)
\bibitem{GB}  M. Bando, T. Kugo, T. Noguchi and K. Yoshioka,
{\it Phys. Rev. Lett.} {\bf 83}, 3601 (1999)     \\
 J. Hisano and N. Okada, {\it Phys. Rev.} {\bf D61}, 106003 (2000)\\
  R. Contino, L. Pilo, R. Rattazzi and A. Strumia, {\it JHEP}
{\bf 0106}:005, (2001)
\bibitem{Kugo} T. Kugo and K. Yoshioka, {\em Nucl. Phys.} {\bf B594}, 301 (2001)
  \\
 P. Creminelli and A. Strumia, {\em Nucl. Phys.} {\bf B596} 125
(2001)
\bibitem{Hamed2} N. Arkani-Hamed, S. Dimopoulos and G. Dvali,
{\it Phys. Rev.} {\bf D59}, 086004 (1999)\\ I. Antoniadis, N.
Arkani-Hamed, S. Dimopoulos and G. Dvali, {\it Phys. Lett.} {\bf
B436} (1998) 257
\\ T. Banks, M. Dine and A.
Nelson, {\it JHEP} {\bf 9906}, 014 (1999)
(1987)
\bibitem{Shifman} G. Dvali, I.I. Kogan and M. Shifman
{\it Phys. Rev.} {\bf D62}, 106001 (2000)
\bibitem{Skyrme}  T.H.R. Skyrme, {\em Proc. Roy. Soc. London}
{\bf 260}
 (1961) 127 \\
 T.H.R. Skyrme, {\em Nucl. Phys.} {\bf 31 }(1962) 556
\bibitem{Witten2} E. Witten, {\it Nucl. Phys.} {\bf B223}, 422 and
433 (1983)
\bibitem{Dobado} A. Dobado and A.L. Maroto
{\it Nucl. Phys.} {\bf B592}, 203 (2001)
\bibitem{Weinberg}  S. Weinberg, {\em Physica} {\bf 96A} (1979) 327\\
   J. Gasser and H. Leutwyler, {\em Ann. of Phys.} {\bf 158}
 (1984) 142
\bibitem{DH}  A. Dobado and M.J. Herrero, {\em Phys. Lett.} {\bf B228}
 (1989) 495 and {\bf B233} (1989) 505
\bibitem{nuestro2} J.A.R. Cembranos, A. Dobado and A.L. Maroto, hep-ph/0106322
\bibitem{Atiyah}  M.F. Atiyah and N.S. Manton, {\em Phys. Lett.} {\bf B222}
\bibitem{Witten1} G.S. Adkins, C.R. Nappi and E. Witten, {\em Nucl. Phys.}
 {\bf B228}  (1983) 552
\bibitem{Clements} M. G. Clements and S.H. Henry Tye,
{\it Phys. Rev.} {\bf D33}, 1424 (1986)\\
A. Dobado and J. Terr\'on,{\em Phys. Lett.} {\bf 247B} (1990)
581
\end{document}